# Scheduling in Data Intensive and Network Aware (DIANA) Grid Environments

Richard McClatchey[1], Ashiq Anjum[1,3], Heinz Stockinger[2], Arshad Ali[3],
Ian Willers[4], Michael Thomas[5]

1) CCS Research Centre, University of the West of England, Bristol, UK
2) Swiss Institute of Bioinformatics, Lausanne, Switzerland
3) National University of Sciences and Technology, Rawalpindi, Pakistan
4) CERN, European Organization for Nuclear Research, Geneva, Switzerland
5) California Institute of Technology, Pasadena, California, USA

**Abstract**

In Grids scheduling decisions are often made on the basis of jobs being either data or computation intensive: in data intensive situations jobs may be pushed to the data and in computation intensive situations data may be pulled to the jobs. This kind of scheduling, in which there is no consideration of network characteristics, can lead to performance degradation in a Grid environment and may result in large processing queues and job execution delays due to site overloads. In this paper we describe a Data Intensive and Network Aware (DIANA) meta-scheduling approach, which takes into account data, processing power and network characteristics when making scheduling decisions across multiple sites. Through a practical implementation on a Grid testbed, we demonstrate that queue and execution times of data-intensive jobs can be significantly improved when we introduce our proposed DIANA scheduler. The basic scheduling decisions are dictated by a weighting factor for each potential target location which is a calculated function of network characteristics, processing cycles and data location and size. The job scheduler provides a global ranking of the computing resources and then selects an optimal one on the basis of this overall access and execution cost. The DIANA approach considers the Grid as a combination of active network elements and takes network characteristics as a first class criterion in the scheduling decision matrix along with computation and data. The scheduler can then make informed decisions by taking into account the changing state of the network, locality and size of the data and the pool of available processing cycles.

**Key words**: meta-scheduling, network awareness, peer-to-peer architectures, data intensive, scheduling algorithm.

## 1. Background

Resource management [1],[2] is a central task in any Grid system. Resources may include "traditional" resources such as compute cycles, network bandwidth, and storage systems. Typical resource management systems that have been proposed are Globus GRAM [3], WMS [4] from EDG [5], GridWay [6], SGE [7], Condor [8] and the EuroGrid-Unicore [9] resource broker projects. Effective resource management and scheduling is a challenging issue, and data location and network load in addition to the computing power are critical factors in making scheduling decisions. The quality and consistency of networks are among the most important factors in this scheduling paradigm since the Grid can be subject to failure if networks do not perform. Moreover, a site that has the targeted data may not be the optimal place for the intended computation even if it has sufficient available computing power since its processors might be required to wait to fetch remote data (and therefore again be dependent on the network load). Similarly, a site with the required data may not be the optimal location to perform the computation if it does not have sufficient available computational resources. All these parameters must be considered in making efficient scheduling decisions.

Grid applications are becoming increasingly network dependent with more demanding requirements in areas such as data access or interactivity. The specific kind of tasks that request computation are usually referred to as "jobs" and are dependent on storage, network capacity and computation. When a job is submitted to a Grid scheduling system, the scheduling system has the responsibility to select a suitable resource and then to manage the job execution. The decision of which resource should be used is the outcome of a matchmaking process between submission requests and available resources. However, during this matchmaking process, we need some adaptive scheduling mechanisms, with appropriate heuristics, which can take into account the characteristics of the network to enable efficient scheduling of data intensive jobs to viable computing resources.

A so-called "meta-scheduler" could facilitate the requesting of resources across multiple machines for jobs and could perform *load balancing* of workloads across multiple sites. Each site also has its own local scheduler to determine how its job queue is processed. A meta-scheduling system works on the basis that a "new task" which needs to be executed has to make itself known to a so-called "matchmaker". This matchmaker acts as a gateway to the Grid. It selects resources from a global directory (e.g. an information service) and allocates the job to one of the available Grid sites.

In this paper, we introduce a Data Intensive and Network Aware (DIANA) meta-scheduling approach which takes into account the network characteristics along with computation and data when scheduling single or bulk jobs. It allows for optimized scheduling decisions as well as reduced execution times of data intensive jobs as documented in our experimental results.

## 2. Problem Description

Data intensive applications analyze large amounts of data which are replicated to geographically distributed sites. If data are not replicated to the site where the job is supposed to be executed, the data need to be fetched from remote sites. This data transfer from other sites will degrade the overall performance of the job execution. In order to support this type of phenomenon, the Grid has a high number of Computing Elements (CE) with Local Resource Management Systems (LRMS) and Storage Services (SE) each identified by a unique ID as shown in Figure 1.

Each data intensive application produces different amounts of data. For performance gains in the overall job execution time and to maximize the Grid throughput, we need to align and co-schedule the computation and the data (the input as well as the output) in such a way that we can reduce the overall computation and data transfer cost. We may even decide to send both the data and executables to a third location depending on the capabilities and characteristics of the computing, network and storage resources. While this might seem to be a simple problem to solve if taken from a local resource management point of view, there is one major issue that must be addressed if we intend to optimize the *overall* Grid throughput at a certain point in time and that is the problem of global distributed scheduling.

Our solution is one of meta-scheduling, a process which allows a user to schedule a job across multiple sites. As we add more complexity to the Grid, particularly with geographically dispersed sites or nodes, it becomes more common for the global meta-scheduler and data mover to make decisions which may be in conflict with decisions that would be arrived at from a local scheduling point of view. This is not surprising since many of the scheduling technologies have been developed under the assumption that all of the collaborating systems are on the same local area network (LAN). If we extend these scheduling or local resource management systems over longer distances, we can begin to see the limitations of the global decision-making systems.

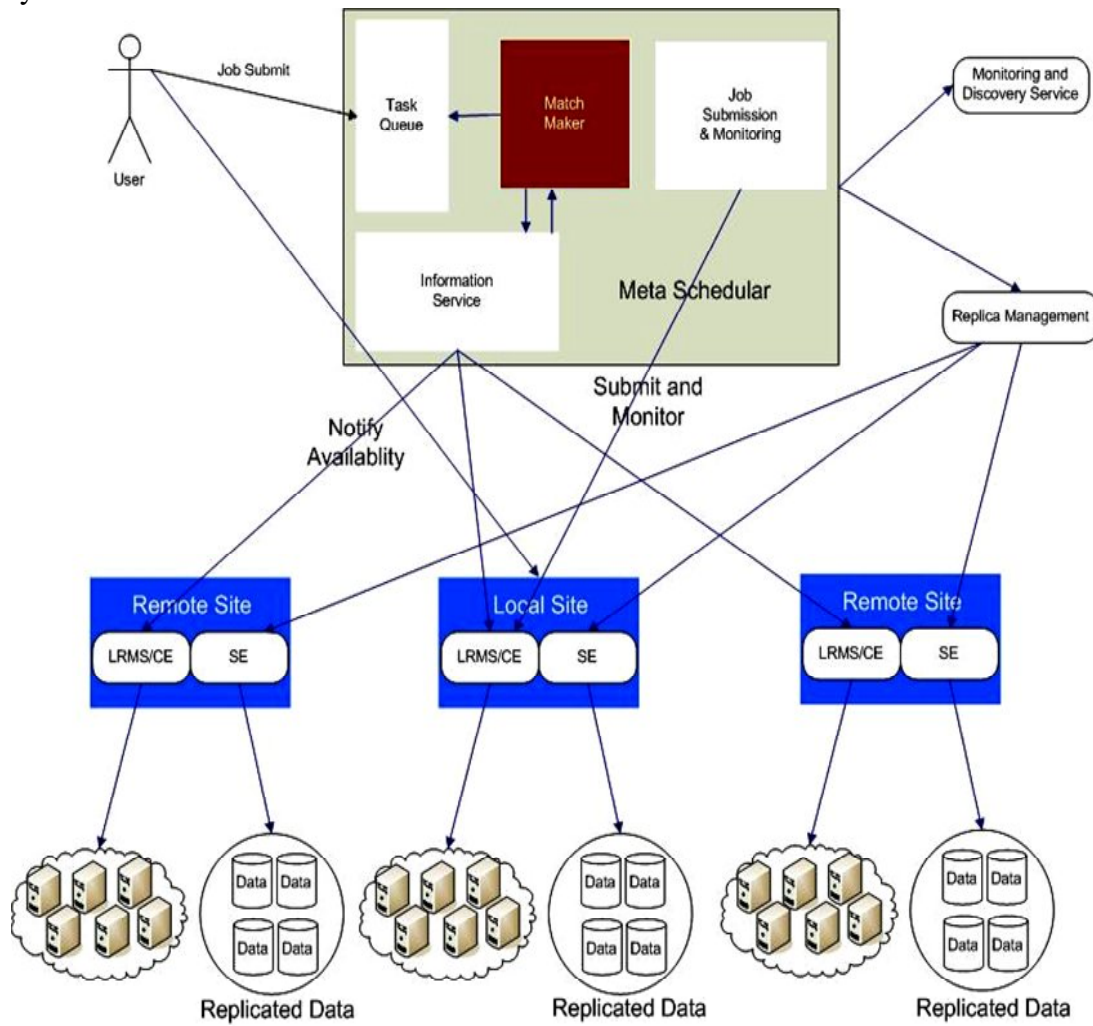

**Figure 1**: Multiple sites Grid meta-scheduling.

*2.1 CMS Physics Analysis Use Case*

Large-scale data-intensive problems, such as those that arise in the high-energy physics (HEP) experiments currently being developed at CERN, can generate petabytes ($10^{15}$ bytes) of scientific data. In HEP data analysis several hundred end-users can run analysis jobs at the same time, processing large amounts of this data (up to several hundred terabytes) replicated over several tens of Grid sites. Data-intensive jobs will clearly need to take data location into account when scheduling these jobs on the HEP Grid resources.

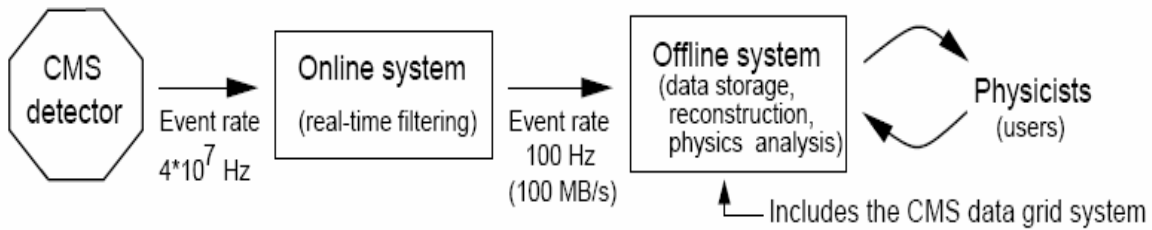

**Figure 2:** The CMS online and offline systems and their environment.

For example, in the CMS experiment at CERN, CMS Production [15] that is the creation of data sets from the experiment itself and CMS Data Analysis [16] are two very data- and computation-intensive processes. CMS Data Analysis deals with the issues of so-called "event reconstruction", the selection of the physics events for further study and the visualization of the data as shown in the Figure 2. In both cases, physicists submit, individually and collectively, millions of jobs (known as bulk submission).

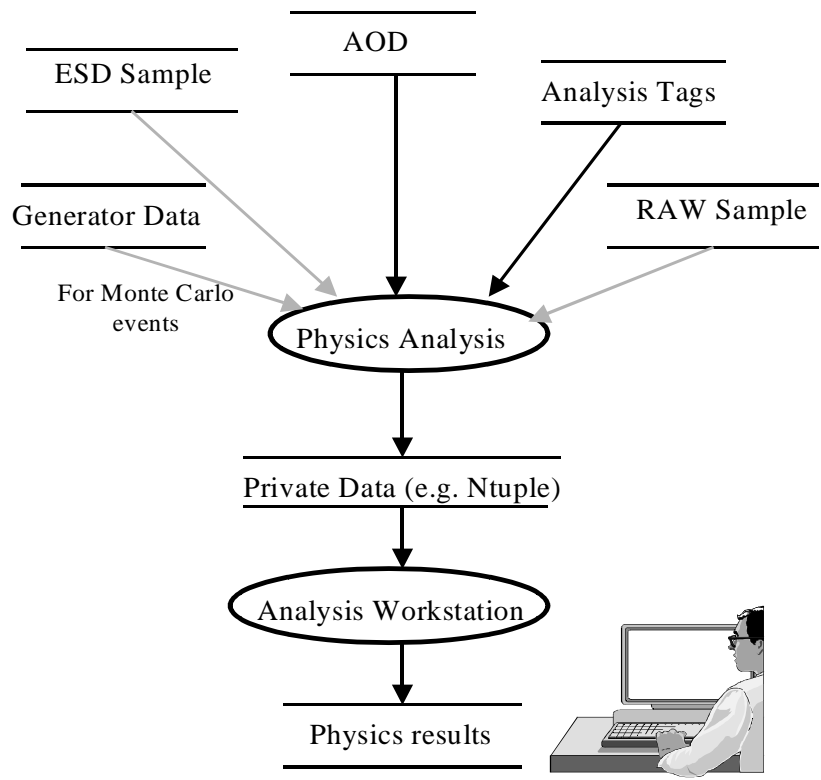

**Figure 3**: User analysis. ESD, AOD and RAW are different data formats

The resource access patterns used in end-user physics analysis (see Figure 3) tend to be less predictable. This comes from the fact that jobs are initiated from almost any CMS site in the world, as well as from the large variation in the "sparseness" of the data access. As shown in Figure 3, when a physicist runs analysis jobs, she may either execute an all-inclusive analysis, using all the collected data, or sub-select interesting events using so-called Tags (i.e. indices to existing data). A typical job will perform some calculation on a specified input dataset and will produce some output. It can be interactive or batch in nature and is part of the dataflow explained above. There are two main cases of CMS jobs:

- Organized jobs. These jobs are planned in advance and perform a homogeneous set of tasks. The input is a pre-determined set of data items (physics events) accessed sequentially, processed and then written out, in a different format, suitable for calculations to be performed in a subsequent analysis phase.
- Chaotic jobs. These jobs are submitted by many users acting more or less independently, and encompass a wide variety of tasks. The input is typically a selection/analysis algorithm to be applied to a very large dataset

This division is in part based on the differences in data-access patterns of the two types of jobs. CMS jobs typically access, process and create large quantities of data, possibly performing nontrivial processing for each event. Jobs of a similar nature are composed in the form of a single bundle and are called the assignments. Currently, a bulk submission of jobs is employed in production and an assignment of more than one job of a similar nature is submitted to a particular site. This is known as Bulk Scheduling. Since an average CMS simulation job (i.e. a typical physics job which creates simulated data) can take around 12-24 hours of computation, in total an assignment may take a week or more to produce the required dataset. A similar scheduling strategy is being planned for CMS physics analysis (i.e. the access to the data generated during the production step), the basic difference being that production jobs produce a large amount of data whereas analysis jobs generally consume data produced by the production jobs.

Presented below are indicative estimates [17] for the typical numbers of jobs from users, together with their computation and data related requirements, which must be supported by the CMS Grid. For each parameter, the first value given is the expected value that needs to be supported, as a minimum, by the Grid system to be useful for the CMS experiment. The second value, between parentheses, is the expected value that is needed to support very high levels of usage by individual physicists:

- Number of simultaneously active users: 100 (1000)
- Number of jobs submitted per day: 250 (10,000)
- Number of jobs being processed in parallel: 50 (1000)
- Job turnaround time: 30 seconds (for tiny jobs) - 1 month (for huge jobs) (0.2 seconds - 5 months)
- Number of datasets that serve as input to a sub job: 0-10 (0-50)
- Average number of datasets accessed by a job: $10^7$ (250,000)
- Average size of the dataset accessed by a job: 1.3 TB (30 GB)

Note that the parameters above have very wide ranges of values; therefore simple averages are not very meaningful in the absence of known variances. Given these statistics regarding workloads, it is clearly challenging to intelligently schedule tasks and to optimize resource usage over the Grid. This has led us to consider a bulk scheduling approach since simple eager or lazy scheduling models are not sufficient for tackling such 'chaotic' analysis scenarios.

*2.2 Wide-Area Scheduling*

A Grid has a number of computational sites which are connected by a number of different WAN links, with different bandwidths and latencies. Consequently, some executables and data items will be "large" compared to the available network bandwidths and latencies. Given these constraints on the overall system, we need to find a way to distribute workloads across all the available systems to maximize the utilization of the available systems. The costs of performing computation will vary according to the types of machines used, the network bandwidth consumed, the state and reliability of the network system, the losses and routing issues in the networks, and will depend on other factors, such as the amount of the data required. These parameters need to be considered when Grid work is scheduled to ensure the effective use of resources in order to minimize cost but maximize workload throughput.

Hence, scheduling algorithms that focus only on maximizing processor utilization by mapping jobs to idle processors and disregard network costs and costs associated with accessing remote data are unlikely to be efficient. Similarly, the scheduling decisions which force the job movement towards the data without taking network load into consideration can lead to significant inefficiencies in performance and can be responsible for large job queues and processing delays.

We not only need to use the network characteristics while aligning data and computations, we also need to optimize the task queues of the meta-scheduler on the basis of this correlation. As a consequence network characteristics can play an important role in the matchmaking process and on Grid scheduling optimization. Therefore a more complex scheduling algorithm is required that should consider the job execution, the data transfer and their relation with various network parameters on multiple sites. Our major challenge then becomes finding a means to express these requirements in a format that the meta-scheduler engine can understand.

The basic job scheduling algorithm needs to be driven by a weighting factor calculated for each potential target location which is a function of the available network characteristics, the processing cycles and data location with the one having least cost being given highest priority. The algorithm should consider the Grid as a combination of the active network elements and must take the network as a first class criterion in the scheduling decision matrix. The scheduling of resources and/or the moving of data from place to place as needed as well as overseeing the task execution through to completion need to be performed on the basis of a "network and strategic view" of the overall Grid system.

## 3. Scheduling Optimization Algorithm

There are three main phases of scheduling on a Grid:
- Phase 1 Resource Discovery: generating a list of potential resources.
- Phase 2 Matchmaking. (This is where this paper makes its main contribution) and
- Phase 3 Job Execution including file staging and cleanup.

In the second phase, the choice of the best combinations of *jobs* and *resources* is a challenging one. We need to embed network information into the scheduling algorithm in order to improve the efficiency and the utilization of a Grid system. The overall goal is to minimize the execution time of applications which involve large scale data.

*3.1 Input Parameters*

The following input parameters are required for our network-aware scheduling and matchmaking algorithm:

- Bandwidth, latencies, packet loss, jitter and anomalies of network links.
- Computing cycles available
- Site loads and respective job queues
- Size of the application executables as well as size of data

*3.2 Optimization Objectives*

The following are the intended objectives within the scheduling process likely to be optimized by this work:

- Queue time and waiting time
- Site load and processing time
- Transfer time for data, executables and results

The total time for job completion and getting results from the job execution in a Grid environment will be the sum of all of these times.

*3.3 Cost Estimators*

There are three major cost estimates which need to be calculated for the scheduling algorithm: network, computation and data transfer cost.

3.3.1 Network Cost

First and foremost is the network cost which depends on many individual parameters. The load, capacity and availability of network links used during data transfers may heavily affect the Grid application performance [25]. Application usage of the network often requires near-real-time, or even real-time, information feedback on the available resources and intelligent decisions on how best to take advantage of these resources [24]. In order to provide the right quality of service (QoS) to Grid applications and hence scheduling, it is important to first understand how the network is performing and to determine the level of quality of service that currently exists in the network. This is measured using four variables, namely *latency*, *dropped packets*, *throughput* and *jitter*.

TCP throughput can be obtained by combining the losses and the Round Trip Times (RTTs) using Mathis's formula [18] for deriving the maximum TCP throughput. Given the historical measurements of the packet loss and RTT, we can calculate the maximum TCP bandwidth for a certain amount of time for various groups of sites. Paper [18] describes a short and useful formula for the upper bound on the transfer rate:

$$\text{Rate} < \left(\frac{MSS}{RTT}\right) \times \left(\frac{1}{\sqrt{loss}}\right)$$

Where:
  *Rate*: is the TCP transfer rate
  *MSS*: is the maximum segment size
  *RTT*: is the round trip time (as measured by TCP)
  loss: is the packet loss rate.

It is clear from the above equation that RTT, TCP throughput or bandwidth and packet loss (including out of order packets and duplicate packets) should be made part of the scheduling algorithm since it has to deal with large data transfers when scheduling data intensive jobs. One way of measuring the quality of service is to measure the number of packets being dropped (the so-called "packet loss").

However, packet loss is not the only cause of poor performance, so care is needed in diagnosing whether genuine packet loss is being experienced. The response time or RTT is the second parameter that can give an idea of the ping data rate (KB/s). The RTT says nothing about how much information a server site can send in a given period. Moreover, for better quality of service and network predictability, we also need to include the jitter in the scheduling algorithm. Overall, as the network utilization increases, the number of dropped packets and the amount of jitter also increases. Consequently, the network cost is the combination of all of the above parameters. We assign weights to each value depending on the importance of the parameters in calculating an aggregate value of the network cost (NetCost).

$$NetCost \propto \frac{Losses}{Bandwidth}$$

*where*

$$Losses = RTT \times W1 + Loss \times W2 + Jitter \times W3$$

Where $W_i$ is the weight assigned to each parameter depending on the importance of the particular parameter. Weights are uniformly assigned subject to a cost; a higher cost indicates the importance of that parameter in the scheduling decision and in some cases we can manipulate these weights to prioritize particular parameters in the algorithm (see Section 3.3.6 for a discussion of the allocation of weights). For example, increasing the weight associated with network cost would bias in favour of data intensive scheduling. To tend towards compute intensive jobs, the compute cost weights can be increased. A higher RTT indicates that a computation site is distant from the storage site where the data resides and therefore the cost to fetch the data would increase. We can increase the significance of this parameter, if required, by assigning a higher value to its associated weight. Higher bandwidth reduces the cost of data transfer and hence the job execution. We can accommodate this behaviour by assigning a higher value to its weight $W_i$. Moreover, jitter is of less importance for data intensive applications and has no significant cost involved due to a higher or lower jitter. We can assign a minimal weight to this parameter but cannot ignore it completely since if this parameter has a value more than an acceptable figure, then there is some bottleneck involved and this element should then have a higher value to reflect this in the overall scheduling algorithm. The same is the true for the packet loss: a higher packet loss implies a less reliable network, and we should give less importance to such a connected site when making scheduling decisions.

There are a number of issues which can influence the scheduling decisions from the network point of view and can lead to skewed results. Selecting the best source from which to copy the data requires a prediction of future end-to-end path characteristics between the destination and each potential source. An accurate prediction of the performance obtainable from each source requires the measurement of available bandwidth (both end-to-end and hop-by-hop), latency, loss and other characteristics which are important in file transfer performance. Because network characteristics are highly dynamic, each reported observation must be attributed with timing information, indicating when the observation was made. Route flaps or other instabilities mean that the same end-to-end traffic may experience a completely different environment from moment to moment.

Frequently in high-bandwidth environments the hosts performing the measurement are the bottleneck, rather than the network path. In the Internet, the path from a source to a destination may be different from the path from the destination back to the source (i.e. asymmetric paths). Even in the case of symmetric paths for both directions, additional traffic from other applications may cause different queuing behaviour in the two directions. Roundtrip measurements can therefore mix the characteristics of both path directions, thereby producing possibly misleading results. Furthermore host timing issues can be problematic for the bandwidth measurement. As the link speeds increase, intra-host latencies cause a greater difference in measurement accuracy. Similarly clock resolution can negatively influence the results since when higher speeds are approached, such as 10 Gbits/s, inter-packet delay timing requires clocks with resolution of 1 µs or better. The "Bulk Transfer Capacity" (BTC) [47] definition of bandwidth measurement assumes an "ideal TCP implementation", which, in practice, does not exist. Since there are many TCP implementations on the Grid, any methodology that relies on a system's TCP implementation is subject to its influence on its results.

3.3.2 Computation Cost

The second important cost which needs to be part of the scheduling algorithm is that of computation cost. Paper [19] describes a mathematical formula to compute the processing time of a job. It is based on Little's theory.

$$\text{Computation Cost} = \frac{Q_i}{P_i} \times W5 + \frac{Q}{P_i} \times W6 + SiteLoad \times W7$$

Where $Q$ is the total number of the waiting jobs on all the sites, $Q_i$ is the length of the waiting queue on the site $i$, $P_i$ is the computing capability of the site $i$ and SiteLoad is the current load on that site. SiteLoad is calculated by dividing the number of jobs in the queue by the processing power of that site. The $Q_i/P_i$ ratio computes the processing time of the job. The $Q_i/P_i$ ratio of the two sites cannot be the same since the number of jobs submitted to the sites will always be different due to differing SiteLoads and other appropriate parameters such as the data transfer cost of the sites. Again $W_5$, $W_6$ and $W_7$ are the weights which can be assigned depending on the importance of the queue and the processing capability. For example, a larger queue makes a site less attractive for job placement so we assign it a higher weight to make the cost higher. Similarly, site load reflects the current load on a site, so again we assign a higher weight if the load on that site is higher.

It is a challenging task to calculate and predict the dynamic nature of the resources and changing loads on the Grid. The load prediction at a site must be dynamic in nature and the least loaded site at one moment can become overloaded the next moment due to bulk submission. Since we use a non pre-emptive mode of execution, once a jobs gets a CPU we cannot abort and move the job to other site. Moreover, Grid resources differ widely in the performance they can deliver to any given application and, because performance fluctuates dynamically due to contention by competing applications, the scheduler must be able to predict the deliverable performance that an application will be able to obtain when it eventually runs.

3.3.3 Data Transfer Cost

The third most important cost aspect in data intensive scheduling is the data transfer cost which includes input data, output data and executables. Reference [20] describes a mathematical technique to calculate the aggregate data transfer time which includes all three parameters. Here, we do not use bandwidth only to calculate the data transfer cost, rather we use the network cost as calculated in Section 3.3.1. We take the case of remote data and different remote execution sites so that the meta-scheduler can consider a worse-case scenario in scheduling.

Data Transfer Cost (DTC) = Input Data Transfer Cost + Output Data Transfer Cost
                        + Executables Transfer Cost    i.e

$$W_8 \times ID \times NC_{(i-j)} + W_9 \times (AD + OD) \times NC_{(local-j)} + W_{10} \times (N_{(j)} \times (ID + AD) + OD) \times NC_{(j)}$$

Where:  ID = Input Data    AD = Application Data    OD = Output Data   and
        NC = Network Cost and *i* and *j* indicate a certain site

Here, we discuss the three different costs for data transfer. Input data transfer cost is the most significant due to expected large data transfers. Higher network cost will increase the data transfer cost and vice-versa, and we use the associated weight to adjust the value according to its importance. The same is the case for the output data since output data needs to be transferred to the location from where the job was submitted. Application data are executables and user code which will be submitted for execution but might be low compared to the input and output data transfer costs.

We can reduce the response time by moving input data from one site to another that has a larger number of processors, since computational capabilities of a remote site without replicated data can be superior to the capabilities of other sites with replicated data. In this scenario, the input data located in site *i* is transferred to site *j* which has sufficient computational capabilities. Also application codes should be transferred from the local site to site *j*. Then the processing is performed in site *j* and the resulting data will be transferred to the local site.

3.3.4 Total Cost

Once we have calculated the cost of each stake holder, the total cost is simply a combination of these individual costs as calculated in Sections 3.3.1, 3.3.2 and 3.3.3 :

Total Cost C = Network Cost + Computation Cost + Data Transfer Cost

The main optimization problem that we want to solve is to calculate the cost of data transfers betweens *sites* (DTC), to minimize the network traffic cost *between the sites* (NTC) and also to minimize the computation cost of a job *within a site*. To simplify the optimisation problem we assume that any given site can have:

- one or many storage resources (Storage Elements, SEs)
- one or many computing resources (Computing Elements, CEs).

Therefore, we are mainly interested in the wide-area network performance rather than specifying all network details within a site. We assume that the local network latency is roughly homogeneous for all nodes (storage or computing) within a site. We can now calculate the cost of the job placement on each site with respect to the submission site. This will be a relative cost since it will always be measured with reference to the user's location on the Grid. Next, we can populate a cost matrix with cost values against each site. In detail, we look at the number of possible sites and calculate the total cost for each pair (site i – site j) and put that into our cost matrix.

We do not take into consideration all the computation and storage sites in this cost matrix since that would require significant effort in calculating the cost of each site against all others in the Grid and the matrix optimization itself requires further research. This approach is not just all-to-all communication. Had it been so i.e. where a site communicates with all peers in the Grid (all-to-all communication), that would have been prohibitively expensive, and the solution would not have been scalable. If we calculate the network, data and compute costs of all sites from each other in a Grid network, then the information collection and decision making costs will become too high. Particularly for large Grid networks, this would no longer be realistic since the information collection would become a more expensive operation than the decision making itself and this aspect must be considered when making scheduling decisions. Instead, we rank the sites on the basis of storage and computation cost and select the best sites (five in this example), which are then used to populate the matrix. It is necessary that these "best" sites have the least cost of all sites in the Grid since this can leads to a minimum time for the overall job execution and ultimately will yield an optimized scheduling and an optimized Grid. Figure 4 shows an example cost matrix giving the overall cost of job submission from one site to all four others in the Grid. $C_{ij}$ is the total cost of a particular site *i* from any other one *j* in the Grid.

|  | Italy | Austria | Switzerland | UK | Japan |
|---|---|---|---|---|---|
| Italy |  | $C_{21} = 50$ | 45 | 60 | 90 |
| Austria | $C_{12} = 58$ |  | 48 | 65 | 72 |
| Switzerland | 64 | 42 |  | 38 | 85 |
| UK | 72 | 65 | 50 |  | 65 |
| Japan | $C_{15} = 70$ | 72 | 85 | 65 |  |

**Figure 4:** The cost matrix for five example sites

Whether a single job is being submitted to the scheduler or bulk job submission is being managed by the meta-scheduler, the cost matrix is equally valid since the cost mechanism will describe the time and cost of each job in the global perspective. Once the cost matrix is populated, we can find the minimum cost of a particular site from all others sites by searching the cost matrix. Once the site with the least cost is selected, the resource broker (i.e. the submission and execution service) will schedule the job on this site. This cost matrix is the core element of the DIANA Scheduler (and its related services that are described in more detail in the next sections) in selecting the optimal site for job execution. The network cost calculated in the algorithm is used to select the best replica of a dataset which will be used as input to the scheduler.

3.3.5 Potential Limitations of the Algorithm

DIANA mainly relies on performance information collected in the past in order to schedule jobs, i.e. to "predict" the future. Therefore, the system relies on the fact that the future is similar to the past. This is a potential problem in all forecasting methods (including the Network Weather Service etc). In addition, since the cost model depends on very detailed and up-to-date monitoring information, DIANA relies very much on the stability, scalability and accuracy of the underlying monitoring system.

Like many meta-schedulers DIANA uses the push approach where jobs are actually pushed (or sent) to computing elements. This approach has the disadvantage that jobs might fail due to configuration errors at the destination sites. Schedulers with pull approaches take care of this problem. Another way to overcome this is to send jobs that constantly monitor the environment for potential errors.

3.3.6 Allocation of Weights

Depending on the nature of the scheduling problem, it may be appropriate to give some of the cases greater weights than others in computing frequency distributions and statistics. The way we do this is to specify that a certain variable contains the *relative* weights for each case and should be considered as a weight variable. The goal of any weight is to prioritize or characterize different variables according to the chosen measure of contribution or influence. Since the objective in weight allocations is to gauge relative weights rather than actual weight values, arbitrary weighting schemes can attach potentially incorrect weights to the component variable.

Let us explain the philosophy of weight allocations through a worked example. This will also demonstrate suitable weights for data intensive applications and how weights are determined in the case of high throughput applications. Let us suppose we have a 100 GB of data located at a site in Japan and that we have jobs that need to access the data for processing and analysis. We also assume that we have only this single copy of the data across the Grid and that every job wherever is submitted will use this data. Via an information service we have determined that a site in Switzerland has the greatest number of computing cycles available for the analysis since it is the least loaded available site and has fewest jobs in its queue. Moreover, there are 8 CPUs available in Japan and 50 in Switzerland, and the bandwidth between these sites is 100 MB/s. The queue size for the site in Japan is 20 whereas that in Switzerland is 2 jobs only. We also assume that the total jobs in the Grid are 1000 at this point in time.

We calculate the siteload by dividing the jobs in the queue by the number of CPUs (assuming CPUs on all the sites have equal processing power). Now there are two options for data analysis: either we should submit the job to the site in Japan where the required data is available or we should transfer the data to the site in Switzerland where the computing capacity is available. The scheduler must decide where this job should be placed so that it has the least overall execution time. We check the compute cost and data transfer cost to enable the decision. The job is data intensive therefore data transfer should get a higher weight relative to others. The scheduler assigns an equal weight to the compute cost on each site. The network cost should be ignored for the time being since we assume that it will remain constant between the two sites. Since we are making a pre-processing decision, we will take input transfer only. We can ignore the executable transfer cost since this data is minimal as compared to input data. RTT and Loss can be ignored since the network seems to be pretty stable. Normally a weight assigned to one variable will be the same for all sites otherwise we can not compare the strengths or weakness of a site. Weights are assigned in the range 1 to 20 where 20 represents a very significant factor and 1 a very insignificant factor. In this example, the site load is a very important factor since it decides how long a job will wait until a job gets a CPU so we have allocated it a maximum value of 20. Further work is required to simulate the exact behaviour when we vary the weights for different variables and check the outcome on the scheduling optimization. Since in this worked example the data transfer cost for Japan is zero, we have assigned a minimal weight of 10 to the same variable for Switzerland site as otherwise it can bias the whole comparison.

| Site | DTC | Compute Cost | Network | Total Cost |
| --- | --- | --- | --- | --- |
| **Japan** | 0 | 10*20/8+5*1000/8+20*20/8 | 0 | 700 |
| **Switzerland** | 10*100*1024/100 | 10* 2/50 +1000*5/50 +20*2/50 | 20/100 | 10341.2 |

**Figure 5:** Weight allocation and cost

From the above calculation in Figure 5, it is clear that the cost for job placement in Switzerland is much higher than the one in Japan and we should send the job towards the data even if better computing cycles are available at the site in the Switzerland. We see that it is the data placement cost which has reduced the chances of selection for the site in Switzerland. Before actually placing the job on a site in Japan, we have to check if there is any other site where the cost combination is smaller. We have realized that although there is no better site than the one in Switzerland in terms of computation power, the scheduler has found a site in UK where the bandwidth is much higher (10 Gbps) than the Japan-Switzerland link, and we should calculate the cost of job placement for this site. The site in the UK has modest computing cycles (only 30) and there are 10 jobs in the queue. Consequently, we need to calculate the cost of job placement for the site in the UK.

| Site | DTC | Compute Cost | Network | Total Cost |
|---|---|---|---|---|
| **Japan** | 0 | 10*20/8+5*1000/8+20*20/8 | 0 | 700 |
| **Switzerland** | 10*100*1024/100 | 10* 2/50 +1000*5/50 +20*2/50 | 20/100 | 10341.2 |
| **UK** | 10*100*1024/10*1024 | 10* 10/30 +1000*5/30 +20*20/30 | 20/10*1024 | **282** |

**Figure 6:** Cost Calculation for best sites

We can see in this example that the better network link has enabled the scheduler to select a site other than those in Switzerland and Japan (cf. Figure 6) and the favoured solution is to move the data and the job towards this site in the UK. Although this site does not have the data and is not as powerful in compute resources as was the case in Switzerland, its job placement cost is still much lower than the other sites and clearly job placement on this site will reduce the overall execution time significantly.

## 4. Architecture

In the following section we describe the job scheduling architecture where our previously introduced optimization algorithm can be applied. We demonstrate how the DIANA Scheduler is used and describe related services that can be used by a Grid job submission service for selecting a suitable execution site for a job.

4.1 General Architecture

The overall architecture of the DIANA Scheduler is shown in Figure 7. It includes a DIANA meta-scheduler with its internal matchmaker. The meta-scheduler uses network information, an optimized replica provided by a Data Location Service (see Section 5) and other information services to make optimal scheduling decisions. The Data Location Service makes use of the Data Location Interface [21] to find the list of the dataset replicas and then uses network statistics to find the "best" replica which is then used by the scheduler. The DIANA scheduling system is implemented as a peer to peer system as discussed in [48]. It should also be noted that an external job submission/execution system needs to be used since the DIANA meta-scheduler only provides scheduling information but does not take care of the actual dispatching/submission of the job to a local resource management system.

The DIANA meta-scheduler mediates between data providers and data requesters. The first step, which is to discover the available resources, is defined *as resource discovery* in Section 3. A resource request consists of a function to be evaluated in the context of a resource. For example, the request "processing power > 2 GHz" will be evaluated by determining if a resource has an attribute called processing power and if so, if the value of this attribute satisfies the condition "Value(processing power) > 2 GHz". If the request can be successfully satisfied, the matchmaker responds with a list of ranked resources. After this, we use the scheduling optimization algorithm to select the best resource and a job is subsequently scheduled to be executed on this resource.

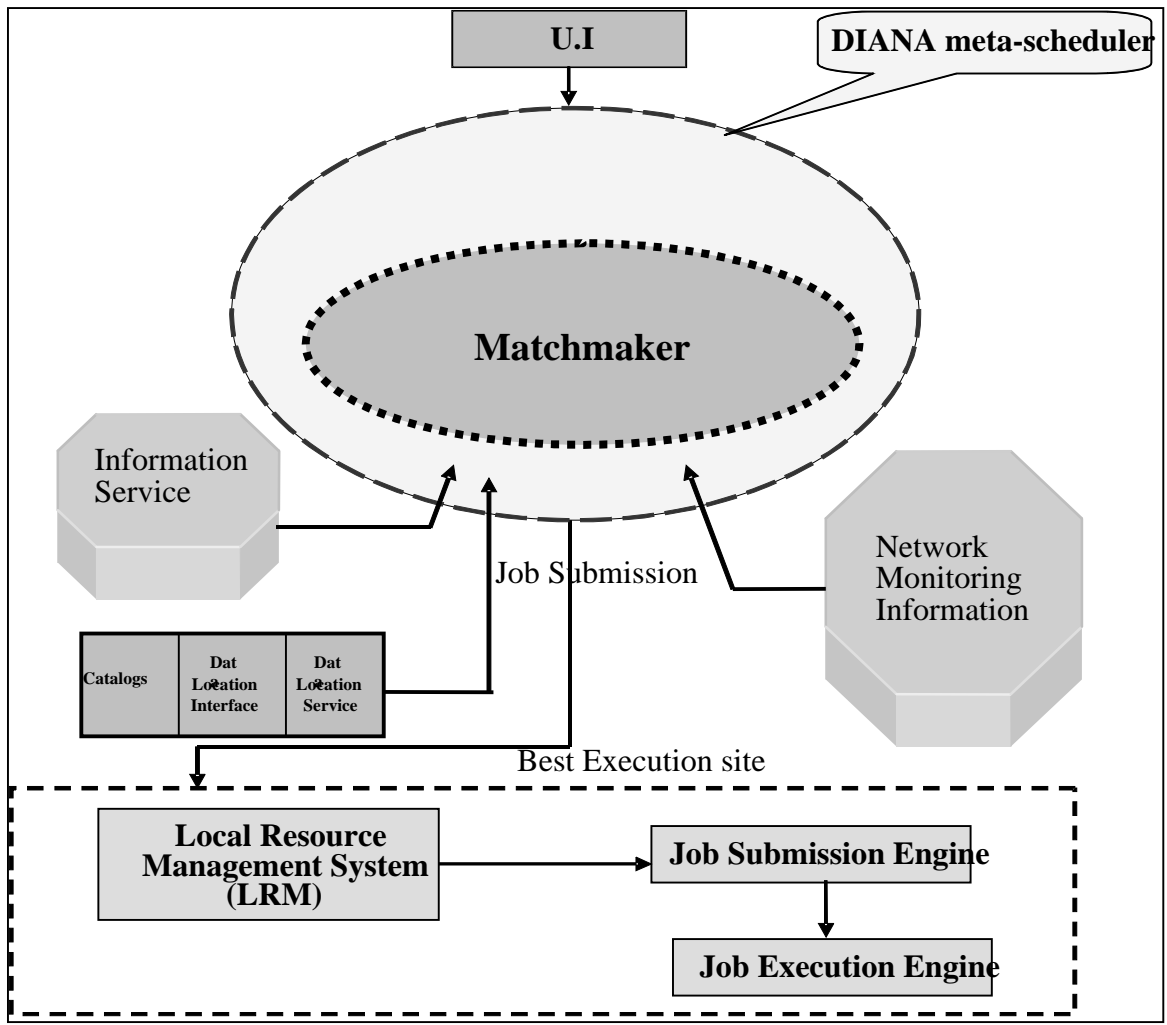

**Figure 7**: A generic job submission architecture with the DIANA Scheduler and the Data Location Service that are used for scheduling data intensive jobs.

The DIANA meta-scheduler keeps track of the load on the sites and selects a site which has a minimum load and queue and has the desired data, processing capability and network stability. Network monitoring information is the central component of the system and all the information collected is stored in a database and is used to make scheduling decisions. The database collects the historical as well as real time information to obtain a current and previous view of the system state.

4.2 The DIANA Scheduler Interface

The DIANA scheduler provides a simple interface for a job submission service (using JDL – Job Description Language [22]) for selecting a suitable Computing Element (CE) for a given data intensive job. The JDL also helps to define the specification of where the data will be read and where the output data needs to be stored. These requirements are passed to appropriate services such as the Data Location Service for decision making. In order to select a certain CE for a specific job with its data requirements, the DIANA Scheduler has two internal methods that work in the following manner:

1) A suitable CE is selected on the basis of computation, network and data transfer cost. The ultimate destination of the output data is also taken into account for selecting the best CE.
2) For the given input data to a job and computing element DIANA finds the "best" replica (represented by a Storage Element, SE), where best refers to the minimal data transfer and network costs. The actual implementation of this functionality is done by the Data Location Service (cf. Section 5).

In more detail, the two methods look like:

- *String GetBestComputingElement ()*
This method returns the best CE with respect to job requirements. This method takes into consideration the number of processors at a site, the load and queue size and the distance from the submission site or from the location where the output data is required. After ranking them, it selects the CE that has lowest cost which is then passed to the *GetBestStorageElement* method below to find an optimal physical replica with respect to that computing element.

- *String GetBestStorageElement (String inputDataType , String inputData , String BestComputingElement)*
This is used to locate the best replica of a dataset. These replicas are ranked with respect to a CE which is selected in the *GetBestComputingElement* method. This CE is passed as a third parameter (BestComputingElement) to this method.

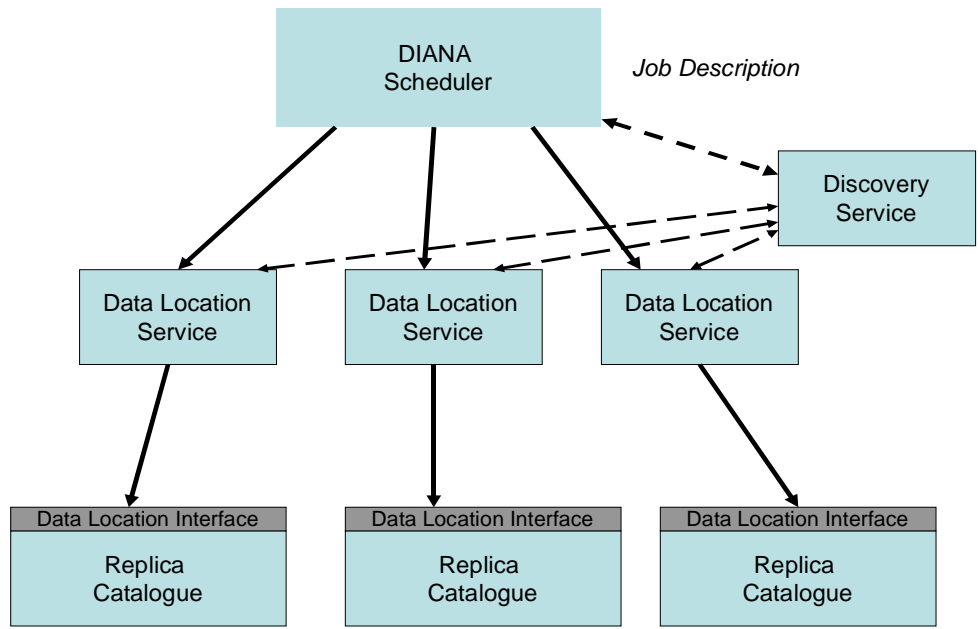

**Figure 8**: Interaction of the DIANA Scheduler with the Data Location Service

## 5. Data Location Service

Figure 8 shows the architecture of the Data Location Service (DLS) implementing the Data Location Interface as described in [21]. In this figure, three instances of the service are shown to illustrate that there can be more than one instance of the DLS running at different locations in the Grid. We can obtain the list of these instances through the Discovery Service which is the point of contact to access and query the DLS. If a client (in our case the DIANA Scheduler) needs to get information about the datasets stored in SEs and registered in replica catalogues recognised by the DLS, the client first contacts the Discovery Service which gives a list of all the locations where a DLS is running. Then the client queries the DLS by passing a logical name of the dataset whose physical location is desired. The DLS is a light-weight Web service that gathers information from the Grid's network monitoring service and performs access optimization calculations based on this information.

The DLS provides optimal replica information on the basis of both faster access and better network performance characteristics. The DLS is fault tolerant, so that when one instance goes offline, a client (service) is still able to work by using another instance of the service. Each of the Grid catalogues is queried by the DLS to find all locations where the requested dataset is available. The service returns a paginated list of dataset locations to the caller. The result of a call to this service is sorted either by the reliability of the datasets which is provided by the network cost and network features such as bandwidth, packet loss etc, or by the "closeness" determined by some network ping time or other network measurements. The DLS also evaluates the network costs for accessing a replica. For this it uses information such as estimates of dataset transfer times based on network monitoring statistics and the replica having the least access and transfer cost is selected.

The Data Location Service uses the DLI [21] to access a replica's information from the various catalogs. A unique feature of the DLI is that it can locate datasets as well as individual files (depending on the underlying replica catalog). A dataset is considered to be an atomic unit of data that is defined within a Virtual Organisation (VO). Furthermore, a dataset itself can consist of several physical files but the end-user (for example a physicist) normally only knows the dataset concept. Due to the distributed nature of the Grid, the files may be replicated at many Grid sites, and the Grid catalogs ensure that the user application does not need to know which locations these are. The Grid catalogs make sure that the file names and associated metadata are properly accessible and secured for the end-user application.

In order to fully include the Data Location Interface into the scheduler, changes were required on the client as well as on the server side. The main change is the integration of the DLI client into the matchmaker. DLI calls follow generally accepted Web service standards.

## 6. Implementation Details

The whole system (that is the DIANA Scheduler with the DLS) is implemented in Java although components and tools employed use C, Python and Perl in addition. We use SOAP as well as XML-RPC for the communication. MonALISA [11] is the core provider of the peer-to-peer (P2P) behaviour in the Discovery Service and it inherits parts of the functionality from JINI. We selected MonALISA since it is the only monitoring tool which can provide the desired P2P behaviour for DIANA. It also provides a suitable Web service interface through which we can integrate DIANA with other Grid services in a loosely coupled way. We have employed PingER [12] to obtain the required network performance information since it provides detailed historical information about the status of the networks. It is a very mature tool that integrates a number of other network performance measurement utilities to provide 'one stop' information for most of the parameters. It does not provide a P2P architecture but information can be published to a MonALISA repository to propagate and access it in a decentralized manner.

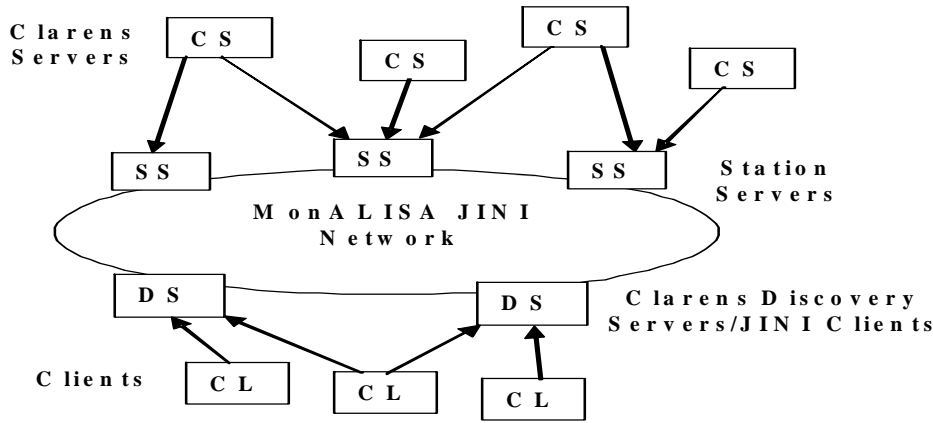

**Figure 9:** Clarens Servers (CS, DS), MonALISA Station Servers (SS), and clients (CL) as part of a P2P discovery system.

DIANA makes use of a peer-to-peer network to track the available resources on the Grid. The current implementation makes use of three software components for resource discovery: Clarens Web services [13],[23] as a resource provider/consumer, MonALISA as a decentralized resource registry, and a peer-to-peer JINI network provided by MonALISA as the information propagation system. The peer to peer behaviour of MonALISA is illustrated in Figure 9 in which it is shown how peers communicate and their fault tolerance capability is demonstrated. The DIANA instances can register with any of the MonaLISA peers through the discovery service and different instances can directly interact with each other.

## 7. Experimental Results

We used the GILDA testbed [26] (a test environment for HEP Grid applications) to validate the results taken by the deployment of the DIANA implementation as discussed in Section 6. The testbed has a series of sites and services (such as a Resource Broker, Information Service, data managers, monitoring tool, CEs and SEs) and is located in several sites in Europe and South America. Figure 10 describes the testbed, its constituent sites, their load and their waiting and running jobs and storage resources on the testbed. The resources are geographically distributed and connected through a high-speed WAN. All of the machines run Scientific Linux CERN. The GILDA testbed has some of the emerging Grid-standard EGEE applications already installed, and we made use of those components and applications. We use the GILDA testbed to run physics analysis applications as a proof-of-concept demonstrator. In addition, we use the EGEE Workload Management System (WMS) to submit and execute jobs that are scheduled by the DIANA Scheduler. Note that we do *not* use EGEE's matchmaker but replace it by DIANA's services.

The main objective of our tests is to reduce (minimize) the *overall job completion time*, i.e. the elapsed wall clock time from submitting the job to the scheduler to actually retrieve finalise the job on the actual computing resource (including writing of output data). We first submitted jobs to the GILDA testbed without the DIANA algorithm of Section 3 and measured the parameters and execution times. After this, jobs are submitted following the algorithm employed in the DIANA Scheduler which includes the measurement of decision parameters as described in Section 3. The total time of job execution is discussed below. Tests are performed by submitting jobs through GILDA's user interface. The client machine was a Pentium based machine with a 2.4 GHz processor and 1 GB RAM The network card was of 100 Mbps capacity. We take a particular CMS computation intensive job which produces a very large amount of data. We selected this CMS job because its execution time is of the order of minutes otherwise as a result of varying network characteristics, there can be a variation in the results in the longer running jobs which can take days.

| Site | | GK# | Q# | RunJob | WaitJob | SlotLoad | MH# | Power | WN# | CPU# | CPULoad | Available | Total | % |
|---|---|---|---|---|---|---|---|---|---|---|---|---|---|---|
| CECUM-ME | 🇮🇹 | 1 | 3 | 0 | 0 | 0% | 2 | 12K | 1 | 2 | 3% | 194.3 GB | 203.3 GB | 4% |
| CENAPAD-UNICAMP | 🇧🇷 | 1 | 3 | 0 | 0 | 0% | 3 | 3K | 1 | 1 | 0% | 59.7 GB | 65.9 GB | 9% |
| CNR-ROMA | 🇮🇹 | 1 | 3 | 0 | 0 | 0% | 6 | 4K | 3 | 3 | 0% | 12.4 GB | 17.2 GB | 28% |
| DIST-GENOVA | 🇮🇹 | 1 | 3 | 0 | 2 | 0% | 5 | 38K | 3 | 8 | 0% | 360.8 GB | 370.5 GB | 3% |
| IHEP-BEIJING | 🇨🇳 | 1 | 3 | 0 | 0 | 0% | - | - | - | - | - | 51.2 GB | 55 GB | 7% |
| IISAS-GILDA | | 1 | 4 | 0 | 3 | 0% | 8 | 24K | 5 | 5 | 20% | 27.2 GB | 28.9 GB | 6% |
| INAF-CATANIA | 🇮🇹 | 1 | 3 | 0 | 0 | 0% | 3 | 3K | 1 | 1 | 0% | 103.1 GB | 106 GB | 3% |
| INFN-CATANIA | 🇮🇹 | 1 | 3 | 0 | 0 | 0% | 13 | 201K | 9 | 36 | 0% | 3 TB | 3.9 TB | 23% |
| INFN-CNAF | 🇮🇹 | 1 | 4 | 0 | 0 | 0% | 4 | 6K | 1 | 2 | 0% | 206.9 GB | 231.4 GB | 11% |
| INFN-PADOVA | 🇮🇹 | 1 | 3 | 3 | 0 | 25% | 8 | 42K | 6 | 12 | 0% | 330.9 GB | 332.9 GB | 1% |
| ING-MESSINA | 🇮🇹 | 1 | 3 | 0 | 0 | 0% | 6 | 36K | 4 | 6 | 0% | 503.6 GB | 521.1 GB | 3% |
| IUCC-LCG2 | 🇮🇱 | 1 | 3 | 0 | 0 | 0% | 7 | 16K | 5 | 10 | 5% | 857.1 GB | 870.1 GB | 2% |
| TRIGRID-UNIPA | 🇮🇹 | - | - | - | - | - | 2 | - | - | - | - | 3 GB | 4.6 GB | 34% |
| prague_cesnet_gilda | - | 1 | 2 | 0 | 0 | - | 5 | - | - | - | - | 2.8 TB | 9.7 TB | 71% |
| **TOTAL** | | #14 | 13 | 40 | 3 | 5 | 2% | 72 | 385K | 39 | 86 | 7% | 8.4 TB | 16.2 TB | 15% |

**Figure 10:** A description of the GILDA Testbed (source https://gilda.ct.infn.it/)

It can be difficult to estimate the true effect of the DIANA scheduling approach if jobs are run at different times, and the results of various approaches are taken at different times. In order to compare the two approaches, we executed short duration jobs in an almost identical environment. DIANA scheduling is equally applicable to short and long duration jobs. For longer jobs it is the execution time which will vary and accordingly queue times will also increase. The *execution cost* will remain the same with time since once the job is submitted, either it is a long or short job, it will not pre-empt until it completes its execution. Therefore, it is not time dependant. The same is the case for the data transfer cost: it should remain the same either a longer job is being executed or a shorter job is being scheduled. The only variable which can change with time is the network cost which can influence the data transfer cost but not the execution times since jobs do not intercommunicate with each other during execution. The data transfer cost is the replication cost and has no link with the job execution time, therefore network cost in DIANA is equally important for longer and shorter jobs. From here we conclude that these results, although are being presented for the short duration job, are equally applicable to long duration jobs and therefore can be generalized.

We submitted a varying number of jobs. First, we submitted 25 jobs and observed their *queue time* and *execution time*. The execution time is the wall clock time taken for a job that is placed on the execution node. It does not include queue time or waiting time. The queue time here is the sum of the time in the meta-scheduler queue and the time spent in the queue of the local resource manager. Then, we submitted the same job three times and measured the queue and execution times again. After this, we increased the number of jobs to 50 and then gradually increased to 1000, so that we can check the capability of the existing matchmaking and scheduling system. We increased the number of the jobs for two reasons. Firstly, to check how the queue size increases and in which proportion the meta-scheduler submits the jobs i.e. whether jobs are submitted to some specific site or on a number of CPUs at different locations depending on the queue size and the computing capability. We calculated and plotted the queue time and investigated how it increased and decreased with the number of jobs.

We observe that both queue and execution time have almost similar trends. This is primarily due to the fact that DIANA selected those sites which can quickly execute jobs (i.e. short local queues with low latency). The queue time is almost proportional to the execution time since if the job is running and taking more time on the processor, the waiting time of the new job will also increase accordingly since it will pass more time in the queue. Although the execution time does not include queue times but a higher number of jobs running at a site can influence the queue time. Furthermore, more jobs in the queue can influence the *overall job completion times* (scheduling time, queuing time and execution time) of the new jobs since they will be competing for the resources to get an execution slot, especially if the jobs are composed of *sub-jobs*. Large jobs are divided into small sub-jobs after a job partitioning process, and most of the time work on the same set of the data. They have similar characteristics and are treated as independent jobs during scheduling, queuing and execution stages. However, their output is returned to the user as a single aggregated unit. These sub-jobs are always scheduled on a single site, and the overall time of the job depends on the execution of these sub-jobs. Some of these jobs will be in the queue and others will be running but the overall time of execution will be the aggregate time when all these sub-jobs complete their execution.

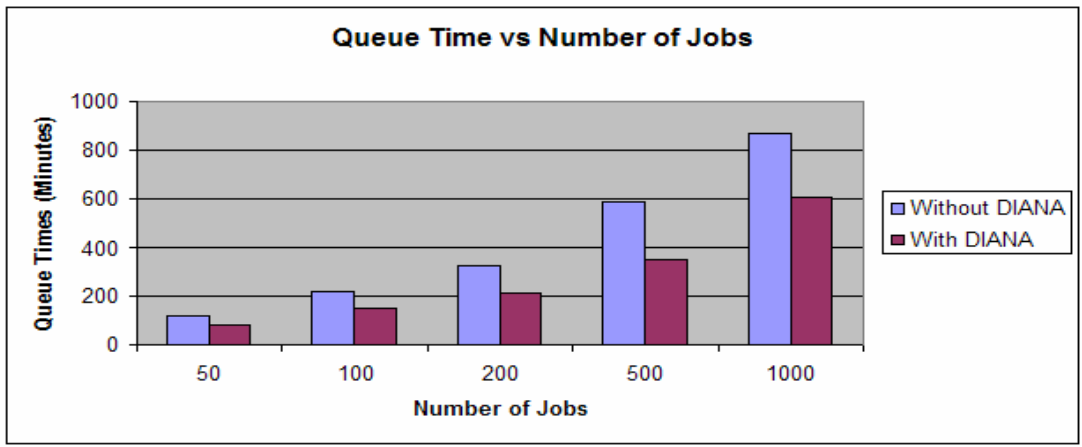

**Figure 11**: Queue time versus number of jobs

The queue time of local resource management systems is very significant in the Grid environment and takes a certain proportion of the job's overall time (cf. Figure 11). Sometimes this is even bigger than the execution time if the resources are scarce compared to the job frequency. We took only a single job queue in the meta-scheduler and we assumed that all jobs have the same priority. In fact, the job allocation algorithm being employed is based on a First Come First Served (FCFS) principle. The FCFS queue is the simplest and incurs almost no system overhead. The graph of the queue times when the number of the jobs changes is shown in Figure 11. It shows that the queue grows with an increasing number of jobs and that the number of jobs waiting for the allocation of the processors for execution also increases. The graph shown in Figure 11 is based on average values of time for varying number of jobs as mentioned before. Improvements in the queue times of the jobs due to DIANA scheduling are also depicted in the same figure.

Similarly, we monitored the *execution times* of the jobs. By increasing the number of the jobs, it is evident from Figure 12 that the overall time to execute a job is increased. As stated earlier this optimization is due to the better selection of the resources, especially in the case of the sub-jobs. This time is calculated by dividing the available computing power by the number of jobs and is indicative of the aggregated execution times. Only one job is executed on a CPU at a time, and jobs can not run in parallel on that CPU since we are following non pre-emptive scheduling model. More CPUs on a site can execute a higher number of (sub-)jobs and more competing jobs clearly mean more time for a specific job to complete. DIANA has improved the execution times of the jobs since it selected only those sites for the job execution which had the required data, less load, fewer jobs in the queue and all this contributed to the execution optimization. Otherwise the sites having a higher number of jobs already running or heavily loaded sites can make the execution times worse.

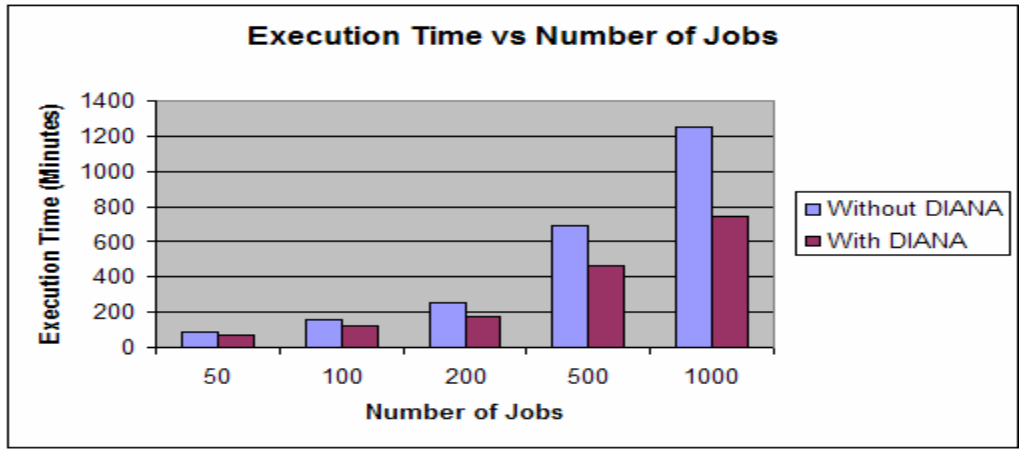

**Figure 12:** Execution time versus number of jobs

Figure 13 (simulation results through MONARC simulations) shows that if the job submission frequency is much higher than the site consumption rate, the site keeps on processing jobs at a constant rate, and the rest of the jobs are exported to other optimally selected sites. The number of jobs that can run simultaneously on a site is equal to the number of the CPUs at that site, and jobs are exported only in exceptional circumstances when the jobs in the queue are estimated to take longer times than the one if exported to a remote site. The export of jobs to other sites depends on the priority of the jobs in a queue and is discussed in [49] which describes a multi-user and priority enabled scheduling mechanism. It is even possible for a site to export jobs which do not have the required data locally.

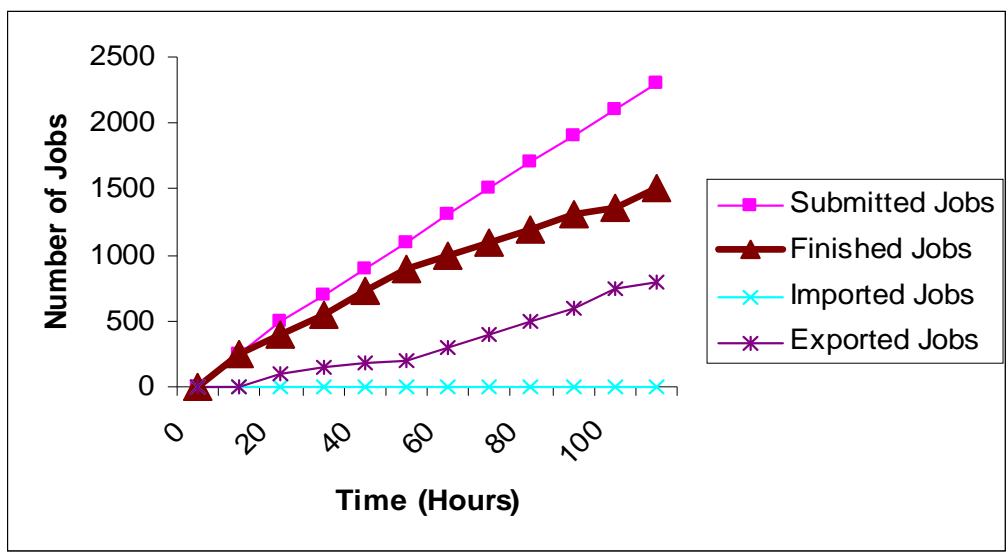

**Figure 13**：The job frequency higher is than the execution capacity of the site

It can be seen that a site can continue processing jobs and at the same time its scheduler can migrate other jobs to more optimal sites according to data availability and job priority. Moreover, a site can simultaneously allow the importation of jobs from other sites which require data that is available on this site or allow the exportation of jobs which can get better execution priority or shorter queues on remote sites. As stated earlier, we employ a non-pre-emptive approach in our scheduling algorithm, and once a job starts execution we do not move it.

Figure 14 shows the replica selection by the Data Location Service which is based on the data transfer cost. We measured the parameters required for calculating the data transfer and network costs. Packet loss and jitter of the sites was almost zero since the network links between the testbed sites are rather stable. Since most of the testbed sites are in Europe, the RTT remains almost the same except the sites in South America and China. Only bandwidth is the parameter which varies across sites and obviously can influence the data transfer and network costs and can dictate the scheduler to select a dataset replica for the job. We took three sets of files of varying size to demonstrate that against each required dataset the scheduler can select a site having the smallest transfer time. From Figure 14 it is clear that for all the three cases the least transfer time is for the site which has a highest bandwidth i.e. 1000 Mbps and most of the jobs use this dataset since it reduces the overall completion time of the jobs. The word "overall" is used since jobs cannot execute until their required data does not become available at the execution site, and jobs keep on waiting in the queue until the data is replicated to the target site. If the data is not replicated and the job is reading data from a remote location, then this will increase the execution time since the job is fetching the data during its execution. Consequently, both queue times and execution times are affected by the replica transfer decisions, and therefore the replica selection contributes to the overall completion time. Obviously, this is valid only for a single user and details can be found in [49] for a multi-user priority enabled scheduling functionality.

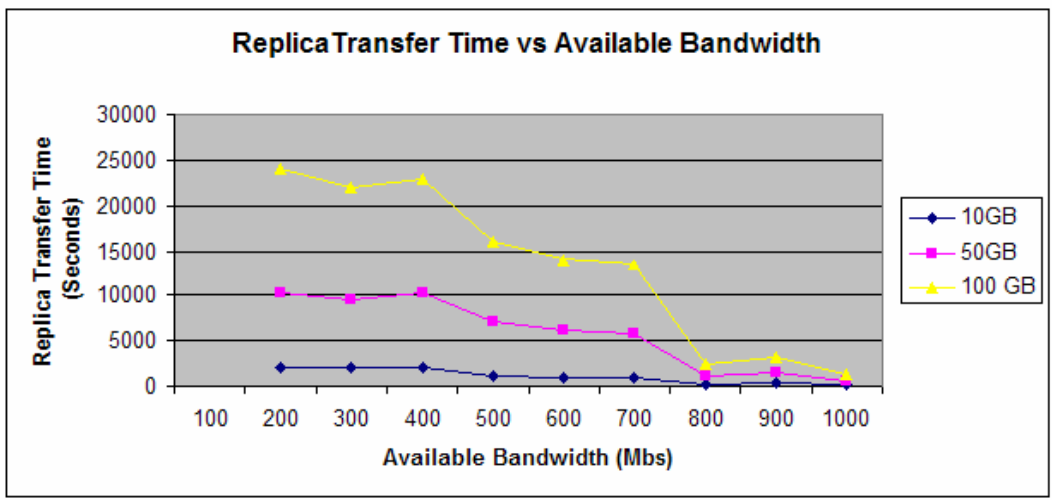

**Figure 14**: Replica Transfer vs. Network Cost

In actual operating conditions for the LHC experiments, there will be hundreds of jobs in the queue as well as in the execution mode, and this means that our scheduling system will make more evident performance improvements than shown in the graphs above. We can ascertain from the graphs that as the number of jobs increases, DIANA has a more profound impact on the scheduling optimization and execution of the jobs. Since most of these jobs take the data from the few selected locations where the replica of that dataset exists, it is assumed that overall completion times will decrease further when thousands of jobs take the actual input data from optimal locations as demonstrated above. In this case, the efficiency of the DIANA scheduling approach will increase further since it is more suited for the environments where huge numbers of jobs are involved and lot of data is taken into account. Consequently, DIANA scheduling helps to decrease the overall completion times of jobs at a given time and therefore provides an efficient way to optimize data intensive Grid jobs.

## 8. Related Work

A number of research projects are under way which are tackling the Grid scheduling issues in general and optimization issues in particular [28]. The most related work that already provided a fully functional Replica Optimization Service (ROS) is described in [29] and [30] and is part of our previous work. The ROS selects the best location for replicas (similar to the DLS described in this article) based on network and storage costs. The system was fully integrated with EDG's workload management system [27] and provided much of the functionality described here. However, the system used a "complicated" network monitoring infrastructure that was not further maintained within the EGEE project and therefore the ROS was no longer deployed. Paper [31] states that hierarchical storage systems are the main source of bottlenecks rather than network parameters but does not consider the meta-scheduling in their findings. The community scheduling framework (CSF) [33] from Platform Computing is a meta-scheduler framework which provides a consistent interface for users into the scheduling system for a Grid. CSF is a centralized engine and it is not intended for bulk data transfer, rather it primarily tackles scheduling heterogeneities. Another similar meta-scheduler is the STAR scheduler [34].

The PhedEX project [35] at CMS is a large-scale data staging, transfer and data scheduling environment. However, there is no concept of job scheduling in PhedEX, and it only does data scheduling for bulk transfers. The Stork project [36] suggests data placement activities are equally important to that of computational jobs in the Grid so that data intensive jobs are automatically queued, scheduled, monitored, managed, and even check-pointed as is done in the Condor project for computation jobs. When combined Condor and Stork do both compute and data scheduling and cover a number of scheduling scenarios and policies, however the functionality of the bulk scheduling has not been considered. Basney et al. [32] define an execution framework which provides an affinity between CPU and data resources in the Grid to run applications on the CPUs which have needed access to datasets but inherent issues and problems in their approach remain the same as those discussed for Condor and Stork. Nathan et al. [14] through their simulation studies suggest a data scheduler for data intensive scheduling but do not give any real implementation of their concept.

The GRESS project [38] is more like a framework in which various algorithms can be plugged in to test their effectiveness. It is only a replication framework and does not provide a scheduling environment. Moreover, Tenet et al. 2005 [39] state performance results of various algorithms and give possible scenarios in which they can increase the scheduling efficiency. They have only evaluated the algorithms but did not present a data intensive scheduling solution. Nabrzyski et al. [40] outline an AI knowledge based meta-scheduler which performs a multi-criteria search technique while making scheduling decisions. Thain et al. [37] describe a system that binds jobs and data together by binding execution and storage sites into I/O communities. The communities then participate in the wide-area system and the Class Ad framework is used to express relationships between stake holders in communities, however the policy issues are not discussed. Their approach does cover co-allocation and co-scheduling problems but does not deal with bulk scheduling and how this can be managed through reservation, priority or policy. The Nimrod-G scheduler [42] works on the principle of deadlines that determine whether jobs are able to complete (in a specified deadline) given the availability of certain resources. It does not include the impact of Grid applications on the system performance. Chamelon [20] implements a data Grid scheduler that takes into account both data location and processor cycles in its decision matrix but their algorithm is based on a 'shortest response time first', and instead we aim at a network aware adaptive algorithm which takes dynamic decisions while scheduling data intensive jobs. The greedy scheduling algorithms used by Chamelon have high resource cost and other shortcomings and this is one reason why we have used network aware adaptive algorithms for our scheduling matrix. SPHINX [43] is a framework for workflow management and execution on heterogeneous platforms and is a data intensive scheduling engine. The AppLes project [44] uses the performance model provided by users to schedule applications. The GrADS [45] project adopts the AppLes scheduling methodology while taking Grid scheduling decisions. Both Apples and GrADS are intended for compute intensive applications and offer very little to accommodate the data intensive or network aware services.

## 9. Conclusions

We presented a Data Intensive and Network Aware (DIANA) scheduling technique in this paper. DIANA takes into consideration data, network and computation power when making scheduling decisions. We proposed a scheduling approach to optimize the completion time of such jobs. We created a scheduling optimization algorithm which takes three different costs and creates a global cost matrix which produces an overall cost for connected Grid sites (and therefore their nodes). The site having the least cost is selected for job execution. We also created a Data Location Service which selects the best physical replica of a dataset.

We implemented the system as a peer-to-peer platform, and our results suggest that there is considerable improvement in the queue as well as execution time of the jobs. We tested the system with a number of jobs, and concluded that system efficiency increases with the number of submitted jobs. This system is efficient in situations where a huge number of jobs are in queues, are data intensive and where the system is complex. The same is the case for the execution time since this scheduling approach makes intelligent decisions about the selection of the execution site, and the final decision is made on the basis of an overall cost. The system is scalable and fault tolerant. Overall results suggest that the scheduling process is optimized as a result of this research, and queue and overall completion times are significantly reduced. Some known problems and potential drawbacks with the system remain the unreliable network conditions which change dynamically and can influence the scheduling decisions. Similarly, the load prediction at a site is also dynamic and the least loaded site at a moment can become overloaded the next moment due to bulk submission. Since we use non pre-emptive mode of execution, once a job gets a CPU, we cannot abort and move the job to other site.

Our future direction is to test the system under various complex scenarios, and we particularly want to test the system on a larger test bed, since GILDA is just an experimental environment. On the other hand, the LHC Computing Grid (LCG) testbed has a huge number of resources, data and many real applications are deployed. These applications will take input data as well as will produce huge amounts of output data and will help us tune the system further. In addition, we also want to conduct some simulation studies of this type of scheduling scenarios so that we can compare the DIANA scheduling approach with the simulated behaviour which should help us further optimize the system. We also plan to explore with simulation/emulation the effect of changing the weights on application performance and want to make weight selection as easy as possible for the user if not automatic.